\documentstyle[pre,aps,multicol,epsf]{revtex}

\tighten

\begin{document}
\draft
\title{Vibronic mechanism of high-$T_{C}$ superconductivity }
\author{M. Tachiki}
\address{ National Institute for Materials Science, 1-2-1 Sengen,\\
Tsukuba, Ibaraki 305-0047, Japan }
\author{M. Machida}
\address{ Center for Promotion of Computational Science and Engineering,\\
Japan Atomic Energy Research Institute, 6-9-3 Higashi-Ueno, Taito-ku\\
Tokyo 110-0015, Japan }
\author{T. Egami}
\address{Laboratory for Research on the Structure of Matter and Department \\
of Material Science and Engineering, University of Pennsylvania,\\
Philadelphia, PA19104, USA\\}
\maketitle

\begin{abstract}
The dispersion of the in-plane Cu-O bond-stretching LO phonon mode in the
high-$T_{C}$ superconducting cuprates shows strong softening with doping near the zone boundary.
 We suggest that it can be described with a negative electronic
 dielectric function that results in overscreening of inter-site Coulomb interaction due to phonon-induced charge transfer and vibronic electron-phonon resonance.
 We propose that such a strong electron-phonon coupling of specific modes can form a basis 
for the phonon mechanism of high-temperature superconductivity. With the Eliashberg theory using the experimentally determined electron dispersion and dielectric function, we demonstrate 
the possibility of superconductivity with the order parameter of the $d_{k_{x}^{2}-k_{y}^{2}}$ symmetry 
and the transition temperature well in excess of 100K.
\end{abstract}


\section{introduction}

The mechanism of the high-$T_c$ superconductivity (HTSC) in the cuprates %
\cite{Bednorz1} remains elusive, in spite of extensive experimental and
theoretical efforts. The majority view on the mechanism is to consider
magnetic interactions as the main driving force \cite{Anderson1}. However,
the HTSC recently observed in MgB$_{2}$ \cite{Nagamatsu1}, graphite-sulfur
composites \cite{Ricardo1}, 
 and the n-type infinite layer
cuprate Sr$_{0.9}$La$_{0.1}$CuO$_{2}$ \cite{Chen} 
cannot be explained by magnetic mechanisms,
since these compounds have little spin fluctuations in strong contrast
to the cuprate superconductors. There is no reason to reject, then, the
possibility that the HTSC in the cuprates also shares a similar non-magnetic
mechanism. In this paper we discuss a phonon mechanism of HTSC based upon
the overscreening of the inter-site Coulomb interaction, in light of recent
experimental results on the cuprates. In this mechanism, unlike 
the BCS theory, the relevant phonons are not the long wave acoustic phonons but
the zone-edge optical phonons that induce inter-site charge transfer.

The HTSC cuprates are doped Mott insulators. Even though the
antiferromagnetism disappears with only about 2 \% of hole doping, strong
antiferromagnetic spin fluctuations are observed by neutron scattering,
nuclear magnetic resonance, and other methods. For this reason spin
fluctuations have been considered to be the principal mechanism of the HTSC %
\cite{Anderson1,Scalapino,Pines1,Moriya1}. However, the intensity of 
spin fluctuations does not correlate, or even anticorrelate, with $T_c$ as the hole
concentration is changed. For instance the spin fluctuations measured by the
nuclear relaxation of $Cu$ in Tl$_{2}$Ba$_{2}$CuO$_{6-\delta }$ are the same
for the sample with $T_c$ = 85 K and the overdoped sample with $T_c$ = 0 %
\cite{Kambe1}. Even when the composition is the same, the spin fluctuations
of La$_{2-x}$Ba$_{x}$CuO$_{4}$ films prepared by epitaxial growth that have $%
T_c$ of 47 K were found to be smaller than those of the bulk with $T_c$ of
30 K \cite{Naito1}.

In this paper we propose a phonon mechanism based upon 
the anomalous screening of the inter-site Coulomb interaction in 
a highly correlated electron system leading to strong pairing. 
This mechanism does not compete against the magnetic mechanism, 
and could achieve HTSC alone or with a magnetic mechanism through 
synergetic effect. There is a large volume of literature that show coupling of 
superconductivity to the lattice and phonon \cite{Pint1,Egami1,Mook1,Lanzara1}. 
In particular the in-plane $Cu-O$ bond-stretching LO phonon mode was observed by 
neutron inelastic scattering to show strong softening with doping near the zone
boundary along the Cu-O bond direction \cite{Pint1,Pint2,McQ1}. This mode
induces charge transfer between Cu and O, and thus couples strongly to the
charge \cite{Egami1,Ishihara1,Petrov1}.  Recent neutron scattering
measurements on YBa$_{2}$Cu$_{3}$O$_{7-\delta }$ (YBCO) suggest that the
frequencies of the LO phonons are strongly softened with doping near the
zone-boundary, while the TO phonons are not \cite{Egami2,Egami3}. We propose
that this occurs due to the phonon-induced charge transfer that results in
formation of a vibronic state \cite{Goodenough} and a negative electronic
dielectric function as will be discussed in section II.

Doped holes in the cuprate superconductors are highly correlated due to the
on-site Coulomb interaction on Cu ions and resultant spin fluctuations. Holes are strongly renormalized due to 
these interactions, and as a result many
physical properties of the cuprate oxides in the normal state show anomalous
behavior. For example, very unusual temperature dependence has been observed 
for the Hall \cite{Hall} and Seebeck coefficients \cite{Seebeck} and the pseude-gap
excitations \cite{Pseudogap} have been observed above $T_c$. The
quasi-particle band structure determined by angular resolved photoemission is 
very different from those calculated with the LDA due to the correlation 
effect \cite{ARPES}. 

Because of these strong correlation effects at present it is very difficult to 
calculate the electronic response function accurately from the first-principles.
 Instead, in the present paper we take a phenomenological approach based upon 
the information obtained from the experimental data. We start with the model 
electronic band structure that agrees with the angle-resolved photoemission 
experiments \cite{ARPES}, and estimate the effective interaction using the phonon 
dispersion determined by the inelastic neutron scattering
 experiments \cite{Egami2,Egami3,Chung1}. 
We then calculate the superconducting order parameter and the superconducting 
transition temperature on the basis of Eliashberg's formulation \cite{Ginzburg1}.
We show that the order parameter is of the $d_{k_{x}^{2}-k_{y}^{2}}$-wave symmetry \cite{d-wave} 
and the superconducting transition temperature can exceed 200K. 
Because of the $d$-symmetry the present phonon-mediated pairing mechanism
does not compete against the spin-fluctuation mechanism. It is possible that 
both mechanisms operate in
the cuprate, and the weight depends upon the doping level. The issue of the
in-plane anisotropy in the phonon dispersion \cite{Egami2,Egami3,Chung1} and
the effect of the possible local spin/charge stripe fluctuation will not be
included in this paper, and will be discussed in future publications.

In section II we explain the idea of overscreening in the cuprate
superconductors. In section III we derive the equation with the charge
kernel for the calculation of the order parameter and $T_c$. In section IV
we determine the energy band renormalized by the correlation effect by
utilizing the experimental results of the angle-resolved photoemission and
the dielectric function using the results of the neutron scattering
measurements. In section V we determine the kernel of the equation by using
the model obtained in Section IV. We then numerically solve the equation and
obtain the superconducting order parameter and estimate the value of the
superconducting transition temperature. The section VI is devoted to
discussions.

\section{Overscreening mechanism and the effective interaction between carriers}

The starting point of this research is to recognize that the electron-phonon
($e-p$) interaction in such a strongly correlated electron system can be
very different from that in the conventional metals. In particular, since
the carrier density in the cuprate superconductors is relatively low and the
charge dynamics is strongly correlated with the spin dynamics, the
medium-range Coulomb interaction is not fully screened. For instance the
optical reflectivity does not saturate as in the standard Drude model even
below the nominal plasma frequency \cite{Optical}. Thus the system has
strong dielectric interactions, unlike in the usual metals. In particular,
since phonons modulate the covalent bonds they induce charge transfer
between ions, and therefore local polarization. In this paper we suggest
that this effect can be described with a negative electronic dielectric
function.

Sometime ago Tachiki and Takahashi proposed an overscreening mechanism of
phonon-mediated superconductivity \cite{Tachiki1,Tachiki2,Tachiki3}. To
explain this mechanism let us consider a static case of the overscreening
effect. We write a staggered external electric field as $D({\bf q})$ and the induced
staggered electric charge polarization as $P_{el}({\bf q})$, and define the electronic charge
susceptibility $\chi_{el}({\bf q},0)$ by 
\begin{equation}
4\pi P_{el}({\bf q})=\chi _{el}({\bf q},0)D({\bf q})
\end{equation}%
Then, combining an electromagnetic relation $D({\bf q})=E({\bf q})+4\pi P_{el} (%
{\bf q})=\epsilon_{el}({\bf q},0)E({\bf q})$ with Eq.(1), we have 
\begin{equation}
\frac{1}{\epsilon_{el} ({\bf q},0)}=1-\chi_{el} ({\bf q},0).
\end{equation}%
According to the Kramers-Kr\"{o}nig relation, we have 
\begin{equation}
\chi_{el} ({\bf q},0)=2\int_{0}^{\infty }d\omega \frac{1}{\omega }\rho_{el} ({\bf q}%
,\omega ),
\end{equation}%
where $\rho_{el} ({\bf q},\omega )$ is the spectral intensity of the electronic charge fluctuation given by 
\begin{equation}
\rho_{el} ({\bf q},\omega )=-\frac{1}{\pi }Im[\frac{1}{\epsilon_{el} ({\bf q},\omega )}]
\end{equation}%
The spectral intensity is always positive and therefore we have an
inequality from Eqs. (2,3,4), 
\begin{equation}
\frac{1}{\epsilon_{el} ({\bf q},0)}\leq 1.
\end{equation}%
The inequality Eq.(5) gives two regions $\epsilon_{el}({\bf q},0)\geq 1$ and $%
\epsilon_{el}({\bf q},0)\leq 0$. The former case is commonly seen for most
materials, such as metals where $\epsilon (0, 0)$ diverges positively,
while the latter case corresponds to overscreening. Let us consider a
positive test charge density $\rho _{test}({\bf q})$. The test charge
induces the screening charge density $\rho _{screen}({\bf q})$ to reduce the
energy of the system. The screening charge is always negative. Using this
charge density, $1/\epsilon_{el} ({\bf q},0)$ is expressed as 
\begin{equation}
\frac{1}{\epsilon_{el} ({\bf q},0)}=1-\frac{|\rho _{screen}({\bf q})|}{\rho
_{test}({\bf q})}
\end{equation}%
When $|\rho _{screen}({\bf q})|$ is larger than $\rho _{test}({\bf q})$ in
Eq.(6), $1/\epsilon_{el} ({\bf q},0)$ is negative. This is the static
overscreening effect.

The microscopic origin of the negative dielectric function may come from the 
Following mechanism.
The electric polarizability in covalent solids is qualitatively
different from that in simple ionic crystals, since the covalency contributes 
to charge transfer between ions \cite{Egami4}. For
instance in ferroelectric oxides the polarization due to charge transfer is
as large as the ionic polarization. In BaTiO$_{3}$ the nominal valence of Ti
is +4 and the nominal $d$ state configuration is $d^0$. But the $d$-orbital
of Ti and the $p$-orbital of O are strongly hybridized. This charge transfer produces
current, and thus electronic polarization, which adds to the ionic
polarization, making the effective valence (Born effective charge) of Ti
twice as large \cite{Resta,Vanderbilt}. In undoped cuprate the same transfer
occurs from the filled $p$-level of O to the empty upper Hubbard band of Cu,
contributory to an extra polarizability.

However, in a doped cuprate the situation is drastically different. The
doped holes occupy mostly the oxygen $p$-levels, and they move either to the 
lower Hubbard band or the filled $d_{z^2}$ orbital of Cu. Thus holes are transferred from O to Cu, creating the
larger polarization, which adds to two ionic polarization.  Then, Eqs. (1,2) give
a negative dielectric function if $\chi_{el}({\bf q},0 )>1$. The more detailed discussions of the lattice softening 
using Born effective charge will be given elsewhere \cite{Piekarz}.

While the in-plane Cu-O bond-stretching LO phonon mode in the cuprates is
strongly softened by doping, other phonon modes are relatively insensitive
to the doping level \cite{Pint1,Pint2,McQ1}. Thus, for the sake of
simplicity, we assume that the basic lattice dynamics is not affected by
doping, and the frequency of the bond-stretching mode is renormalized only
by the interaction with charge fluctuations. This assumption allows to write the frequency corresponding to 
the maximum value of the spectral intensity of the renormalized LO phonon in terms of 
the electronic dielectric function as \cite{Tachiki1,Tachiki2},

\begin{equation}
{\omega _{LO}^{\ast }({\bf q})}^{2}={\omega _{TO}({\bf q})}^{2}+\frac{{%
\omega _{LO}({\bf q})}^{2}-{\omega _{TO}({\bf q})}^{2}}{\epsilon _{el}^{\prime}(%
{\bf q},{\omega _{LO}^{\ast }({\bf q})})},
\end{equation}%
where $\omega _{TO}$ and $\omega _{LO}$ are respectively the bare TO and LO
phonon frequencies in the insulating state, and $\epsilon _{el}^{\prime}({\bf q}%
,\omega _{LO}^{\ast })$ is the real part of the electronic dielectric
function. The experimental results that 
the frequency ${\omega _{LO}^{\ast }({\bf q})}^{2}$ of the LO mode is lower
than that of the TO mode. It indicates that the real part of ${\epsilon _{el}(%
{\bf q},{\omega _{LO}^{\ast }({\bf q})})}$ is negative.
 In YBa$_{2}$Cu$_{3}$O$_{6.95}$ (YBCO) and La$%
_{1.85}$Sr$_{0.15}$CuO$_{4}$ (LSCO) the inversion of the LO/TO frequencies
is most pronounced in the region $q_{x}$ = 0.25 to 0.75 and $q_{y}$ = -0.2 to
0.2, in the units of $2\pi /a$, at $\omega_{LO}^{\ast }$ being the measured
LO frequency (approximately 55 meV for YBCO and 70 meV for LSCO) \cite%
{McQ1,Egami2,Egami3}. We interpret this phenomenon as the consequence of the
negative $\epsilon _{el}^{\prime}({\bf q},\omega _{LO}^{\ast }({\bf q}))$.

The quasi-particles are renormalized due to the on-site Coulomb interaction
as mentioned in Introduction. The interaction increases the effective mass
of the quasi-particles, but does not change the charge ${\it e}$ of the particles.
Therefore, the effective potential acting between quasi-particles 
${\bf k}$ and ${\bf k^{\prime }}$ is written as
\begin{equation}
V_{eff}({\bf q},\omega )=\frac{V({\bf q})}{\epsilon ({\bf q},\omega )},
\end{equation}%
where $V({\bf q})$ is the bare Coulomb interaction and 
${\bf q}$ is ${\bf k}-{\bf k^{\prime }}$. Therefore with the normal screening 
the effective potential is always smaller
than the bare potential. However, in the case of overscreening ${\epsilon_{el }^{\prime}
({\bf q},\omega )}$ can be negative, and thus ${\epsilon_{el }^{\prime}
({\bf q},\omega )}$ contributes to the effective interaction and works to enhance 
the phonon mediated attractive interaction as seen in section III. 
Since the overscreening effect comes from various kinds of the correlation effect, 
${\epsilon ({\bf q},\omega )}$ is a complicate function of charge, spin, and lattice.
This effect is the core of the present mechanism. 

Consequently the attractive interaction should exist between the
quasi-particles with ${\bf k}$ and ${\bf k^{\prime }}$ when ${\bf q}={\bf k}-%
{\bf k^{\prime }}$ is in the ${\bf q}$ regions where the phonon softening
occurs, as seen in Eq.(7). In addition the negative electronic dielectric function
turns the repulsive electron-electron Coulomb interaction into attraction,
as in the so-called negative-U mechanism, and can promote pairing.

\section{ Formulation for Superconducting Pairing}

In this section we derive the equation to calculate the symmetry of the
superconducting order parameter and $T_c$, using the effective interaction
Eq.(8) \cite{Tachiki1,Tachiki2}. In the effective interaction, the dynamical
dielectric function $\epsilon ({\bf q},\omega )$ is given by the sum of the
electronic dielectric function $\epsilon _{el}({\bf q},\omega )$ and the
ionic dielectric function $\epsilon _{ion}({\bf q},\omega )$ as 
\begin{equation}
\epsilon ({\bf q},\omega )=\epsilon _{el}({\bf q},\omega )+\epsilon _{ion}(%
{\bf q},\omega )-1
\end{equation}%
A minus unity in the right hand side of Eq.(9) comes from the fact that all
the dielectric functions should be unity at the high frequency limit. We
express the ionic dielectric function on a conventional form, 
\begin{equation}
\epsilon _{ion}({\bf q},\omega )=\frac{\omega ^{2}-\omega _{LO}^{2}}{\omega
^{2}-\omega _{TO}^{2}},
\end{equation}%
$\omega _{LO}$ and $\omega _{TO}$ being respectively the frequencies of the
bare longitudinal and transverse optical phonons in the insulating state.
For simplicity, we consider one optical phonon mode, which seems to be most
relevant to the superconductivity, and assume that it is dispersionless 
in the unrenormalized state. The spectral intensity function
of total charge fluctuations is expressed by using Eqs.(4),(9-10) as 
\begin{eqnarray}
\rho ({\bf q},\omega ) &=&-\frac{1}{\pi }Im[\frac{1}{\epsilon ({\bf q}%
,\omega )}]  \nonumber \\
&=&-\frac{1}{\pi }Im[\frac{1}{\epsilon _{el}({\bf q},\omega )}]+\frac{\omega
_{LO}^{2}({\bf q})-\omega _{TO}^{2}}{{\epsilon _{el}({\bf q},\omega
_{LO}^{\ast })}^{2}}\delta (\omega ^{2}-\omega _{LO}^{\ast 2}),
\end{eqnarray}%
where the first term in the right hand side of Eq.(11) is the electronic
spectral intensity $\rho _{el}({\bf q},\omega )$, and $\omega _{LO}^{\ast }(%
{\bf q})$ is the LO phonon frequency renormalized by charge fluctuations and
is given by Eq.(1). If we use the spectral representation for $1/\epsilon (%
{\bf q},\omega )$, the effective interaction Eq.(8) is written as 
\begin{equation}
V_{eff}({\bf q},\omega )=V({\bf q})/\epsilon ({\bf q},\omega )=V({\bf q}%
)[1-2\int_{0}^{\infty }d\Omega \frac{\Omega \rho ({\bf q},\Omega )}{\Omega
^{2}-{(\omega +i\delta )}^{2}}],
\end{equation}%
Using Eq.(12) we set up the Eliashberg equation linearized with respect to
the gap function $\Delta ({\bf k},i\omega )$ as 
\begin{equation}
\Delta ({\bf k},i\omega _{n})=-T\sum_{\ell }\sum_{{\bf k^{\prime }}}V_{eff}(%
{\bf k}-{\bf k^{\prime }},i\omega _{n}-i\omega _{\ell })\frac{\Delta ({\bf %
k^{\prime }},\omega _{\ell })}{\xi _{{\bf k^{\prime }}}^{2}+\omega _{\ell
}^{2}},
\end{equation}%
where $\omega \equiv (2n+1)\pi T$ with $n$ being integer, and $\xi _{k}$ is
the quasi-particle energy measured from the Fermi level. We use an
approximation that the damping of the quasi-particles is neglected. However,
the modification of the band structure due to the correlation effect is
taken into account by using the model band structure determined by the
experimental results of angle-resolved photoemmision. If we introduce the
pair function defined by 
\begin{equation}
F({\bf k},i\omega _{n})=\Delta ({\bf k},i\omega _{n})/(\omega _{n}^{2}+\xi _{%
{\bf k}}^{2}),
\end{equation}%
Eq.(13) is rewritten as 
\begin{equation}
F({\bf k},i\omega _{n})=-\frac{1}{(\omega _{n}^{2}+\xi _{{\bf k}}^{2})}%
T\sum_{\ell }\sum_{{\bf k^{\prime }}}V_{eff}({\bf k}-{\bf k^{\prime }}%
,i\omega _{n}-i\omega _{\ell })F({\bf k^{\prime }},i\omega _{\ell }).
\end{equation}%
We then introduce again a function $f(k,v)$ \cite{Russia} defined by 
\begin{equation}
F({\bf k},i\omega _{n})=2\int_{0}^{\infty }dv\frac{vf({\bf k},v)}{\omega
_{n}^{2}+v^{2}},
\end{equation}%
and also 
\begin{equation}
\Phi ({\bf k})=2|\xi _{{\bf k}}|\int_{0}^{\infty }dvf({\bf k},v).
\end{equation}%
Then, from Eqs.(14) and (16) we can show that $\Phi ({\bf k})$ is equal to $%
Re\Delta ({\bf k},\xi _{k})$ to a good approximation. The equation for $\Phi
({\bf k})$ is obtained substituting Eqs.(16) and (17) for Eq.(15) as 
\begin{equation}
\Phi ({\bf k})=-\sum_{{\bf k}}K({\bf k},{\bf k^{\prime }})\frac{tanh\xi _{%
{\bf k^{\prime }}}/2T}{2\xi _{{\bf k^{\prime }}}}\Phi ({\bf k^{\prime }}),
\end{equation}%
In Eq.(18), the kernel $K({\bf k},{\bf k^{\prime }})$ is given by 
\begin{equation}
K({\bf k},{\bf k^{\prime }})=v({\bf k}-{\bf k^{\prime }})[1-2\int_{0}^{%
\infty }d\Omega \frac{\rho ({\bf k}-{\bf k^{\prime }},\Omega )}{\Omega +|\xi
_{{\bf k}}|+|\xi _{{\bf k^{\prime }}}|}].
\end{equation}%
If we insert Eq.(11) for $\rho ({\bf q},\omega )$ in Eq.(19) and use the
Kramers-Kronig relation for $Im(1/\epsilon _{el}({\bf q},\omega ))$ in $\rho
_{el}({\bf q},\omega )$, the kernel Eq.(19) is written as 
\begin{equation}
K({\bf k},{\bf k^{\prime }})=v({\bf k}-{\bf k^{\prime }})[\frac{1}{\epsilon
_{el}({\bf q},0)}+2\int_{0}^{\infty }d\Omega \frac{|\xi _{{\bf k}}|+|\xi _{%
{\bf k^{\prime }}}|}{\Omega (\Omega +|\xi _{{\bf k}}|+|\xi _{{\bf k^{\prime }%
}}|)}\rho _{el}({\bf q},\Omega )-\frac{1}{\epsilon _{el}({\bf q},\omega
_{LO}^{\ast }({\bf q}))}\frac{\omega _{LO}^{\ast 2}({\bf q})-\omega _{TO}^{2}%
}{\omega _{LO}^{\ast }({\bf q})[\omega_{LO}^{\ast }({\bf q})+|\xi _{k}|+|\xi
_{k^{\prime }}|]}].
\end{equation}%
Roughly speaking, in the right hand side of Eq.(20) the first two terms are
the electronic contribution and the third term is the phonon contribution.
However, the electronic contribution is mixed also in the third term as seen
in Eq.(20). Therefore, the kernel has a vibronic nature. 

The spectral intensity $\rho({\bf q}, \omega)$ is approximately written around 
$\omega$ being $\omega_{LO}^{\ast}({\bf q})$ \cite{Tachiki1} as 
\begin{equation}
\rho ({\bf q},\omega )=\frac{1}{\epsilon _{el}^{\prime}({\bf q},\omega_{LO}^{\ast }({\bf q})) ^{2}}
\frac{\omega _{LO}^{2}-\omega _{TO}^{2}}{2\pi \omega _{LO}^{\ast}({\bf q})}
\frac{\Delta(\bf q)}{[\omega-\omega_{LO}^{\ast }({\bf q})] ^{2}+\Delta^{2}({\bf q})}
\end{equation}
with
\begin{equation}
\Delta(\bf q)=\frac{\epsilon _{el}^{\prime\prime}({\bf q},\omega_{LO}^{\ast})}
{\epsilon _{el}^{\prime}({\bf q},\omega_{LO}^{\ast }({\bf q})) ^{2}}
\frac{\omega _{LO}^{2}({\bf q})-\omega _{TO}^{2}({\bf q})}{2\omega_{LO}^{\ast}({\bf q})} 
\end{equation}
In Eq.(22), $\epsilon _{el}^{\prime\prime}({\bf q},\omega_{LO}^{\ast})$ is the imaginary part of 
the electronic dielectric function. $\Delta(\bf q)$ is the half width in the frequency $\omega_{LO}^{\ast}$ of the spectral function. 
If we use the experimental frequency values in Fig.2, $\epsilon _{el}^{\prime}
({\bf q},\omega_{LO}^{\ast})$ is given -0.068 from Eq. (7) for YBCO. Using this value, 
Eq. (22) and the neutron scattering experimental value of $\Delta(\bf q) \sim$ 2meV, 
we obtain $\epsilon _{el}^{\prime\prime}({\bf q},\omega_{LO}^{\ast})$ to be 0.0069. 
Therefore, we can approximate $\epsilon_{el}({\bf q},\omega_{LO}^{\ast}({\bf q}))$ by 
$\epsilon _{el}^{\prime}({\bf q},\omega_{LO}^{\ast})$, while we can see 
that the damping on the low frequency charge fluctuation is very weak.  

In conventional systems with $\epsilon _{el}^{\prime}({\bf q},\omega )\geq 1$ the third term in 
Eq.(20) cannot be enhanced. However, when $\epsilon _{el}^{\prime}({\bf q},\omega)<0$ 
and its absolute value is small, the phonon contribution can be
strongly enhanced, as seen from Eq.(20). The strong softening of 
$\omega_{LO}^{\ast}({\bf q })$ also enhances the third term.

\section {Quasi-particle Band Structure and Electronic Dielectric Function in
Cuprate Superconductors}

In this section, the quasi-particle band structure and the dielectric function are 
determined by utilizing the experimental data of the angle-resolved
photoemission and inelastic neutron scattering.  The results will be used 
in the following section in calculating the symmetry of the superconducting
 order parameter and the superconducting transition temperature with the Eliashberg equation.

\subsection{ Quasi-particle Band Structure}

We assume that the layers responsible for the high $T_c$ superconductivity are
mainly the CuO$_{2}$ layers. Then, we construct the energy band of the CuO$%
_{2}$ layer (the $a-b$ plane) as to agree with the results of the
angle-resolved photoemission spectroscopy (ARPES) \cite{ARPES,Shen2}. The
band structure is almost universal for all the cuprate superconductors with
optimum doping \cite{ARPES}. The angle-resolved photoemission \cite{Shen2}
and the calculation including the correlation effect \cite{Flat} shows the
energy band in the cuprate superconductors is very flat in the region of $%
k_{x}(k_{y})= 0.25-0.75$ in units of $2\pi /a$, at $10-30$ meV below the Fermi
level. Hereafter, we express the wave number in units of $2\pi /a$. This is
the region where the superconducting gap is the largest, and the kink in
dispersion, which is interpreted as an evidence of strong electron-phonon
coupling \cite{Lanzara1}, is most conspicuous. Therefore it is reasonable to
expect that the charge fluctuation created by quasiparticles on the flat
band is very low in energy, comparable to the optical phonon frequency.
Then, the quasi-particles and phonons can be highly mixed, and this system
consequently becomes vibronic. As a result the quasiparticle-phonon
interaction is highly enhanced, as will be shown in sections III and V.

The band structure with the flat regions along the $k_{x}$ and $k_{y}$ axes
which originate from the strong renormalization effect \cite{Flat} is
reproduced by the following function, 
\begin{equation}
\xi
_{k}=C_{B}(t_{0}+2t_{1}(cos 2\pi k_{x}+cos 2\pi k_{y})+4t_{2}cos 2\pi k_{x}cos 2\pi k_{y}+
2t_{3}(cos4\pi k_{x}+cos4\pi k_{y}))+F_{B}(E_{f}-\omega _{g})-E_{f},
\end{equation}%
where $t_{0}$, $t_{1}$, $t_{2}$, and $t_{3}$ are respectively $0.0$, $-0.2$, $0.0$ and $-0.04$, and $E_{f}$
is $-0.11$ and $\omega _{g}$ is from 0.01 to 0.03 in units of eV. In this
equation, if the second term is dropped, the equation approximately
coincides with that of the usual LDA band calculation \cite{Hamada}. The
second term is required to reproduce the flat regions mentioned above. The
function $F_{B}$ is given by 
\begin{equation}
F_{B}=W_{x}( 1+\frac{1}{2}tanh((k_{x}-k_{L})/\lambda _{L})-\frac{1}{2}tanh((k_{x}+k_{L})/\lambda
_{L}))+W_{y}(1+\frac{1}{2}tanh((k_{y}-k_{L})/\lambda _{L})-\frac{1}{2}tanh((k_{y}+k_{L})/\lambda
_{L})),
\end{equation}%
with 
\begin{eqnarray}
W_{x} &=&-\frac{1}{2}tanh((k_{y}-k_{T})/\lambda _{T})+\frac{1}{2}tanh((k_{y}+k_{T})/\lambda _{T}),
\\
W_{y} &=&-\frac{1}{2}tanh((k_{y}-k_{T})/\lambda _{T})+\frac{1}{2}tanh((k_{y}+k_{T})/\lambda _{T}),
\end{eqnarray}%
where $k_{L}$, $k_{T}$ and $\lambda _{L}=\lambda _{T}$ are the parameters to
control the extent of the flat regions. A function $F_{B}$ has sizable
values only in certain regions near the $k_{x}$ and $k_{y}$ axes. For $C_{B}$
and $F_{B}$, a constraint $C_{B}+F_{B}=1$ is imposed. Fig. 1 shows the band
structure calculated by using Eqs.(23-26) where $k_{L}$, $k_{T}$, and $%
\lambda _{L}(=\lambda _{T})$ are, respectively, taken as 0.16, 0.15, and
0.2 in units of $2\pi /a$. The values above are chosen to reproduce the
ARPES data for the optimally doped YBCO.

As seen in the inset (a) of Fig. 1 the flat regions of the energy bands are
located along the $k_x$ axis and extends from (-0.5, 0) to (-0.25, 0) and
from (0.25, 0) to (0.50, 0), and similarly along the $k_y$ axis. The
behavior has been observed in the optimally doped YBCO and LSCO by ARPES %
\cite{ARPES,Shen2}. The energy of the flat band is from 10 meV to 30 meV
below the Fermi level \cite{ARPES}. In Fig. 1, $\omega_g$ is chosen to be 18
meV below the Fermi level. These flat regions mainly originate from the
correlation effect reflecting the electron scattering due to spin
fluctuations \cite{Flat}. In these regions the effective masses and the
damping constants of the quasi-particles are very large. On the other hand,
the energy band in the directions from the (0, 0) point to the (0.5, 0.5),
(0.5, -0.5), (-0.5, 0.5), (-0.5, -0.5) points is not much affected by the
correlation effects and the dispersion of the electronic energy band
structure is usual as seen in the inset (b) of Fig. 1.

\begin{figure}[tbp]
\centerline{\epsfxsize=16.0cm \epsfysize=17.0cm 
\epsfbox{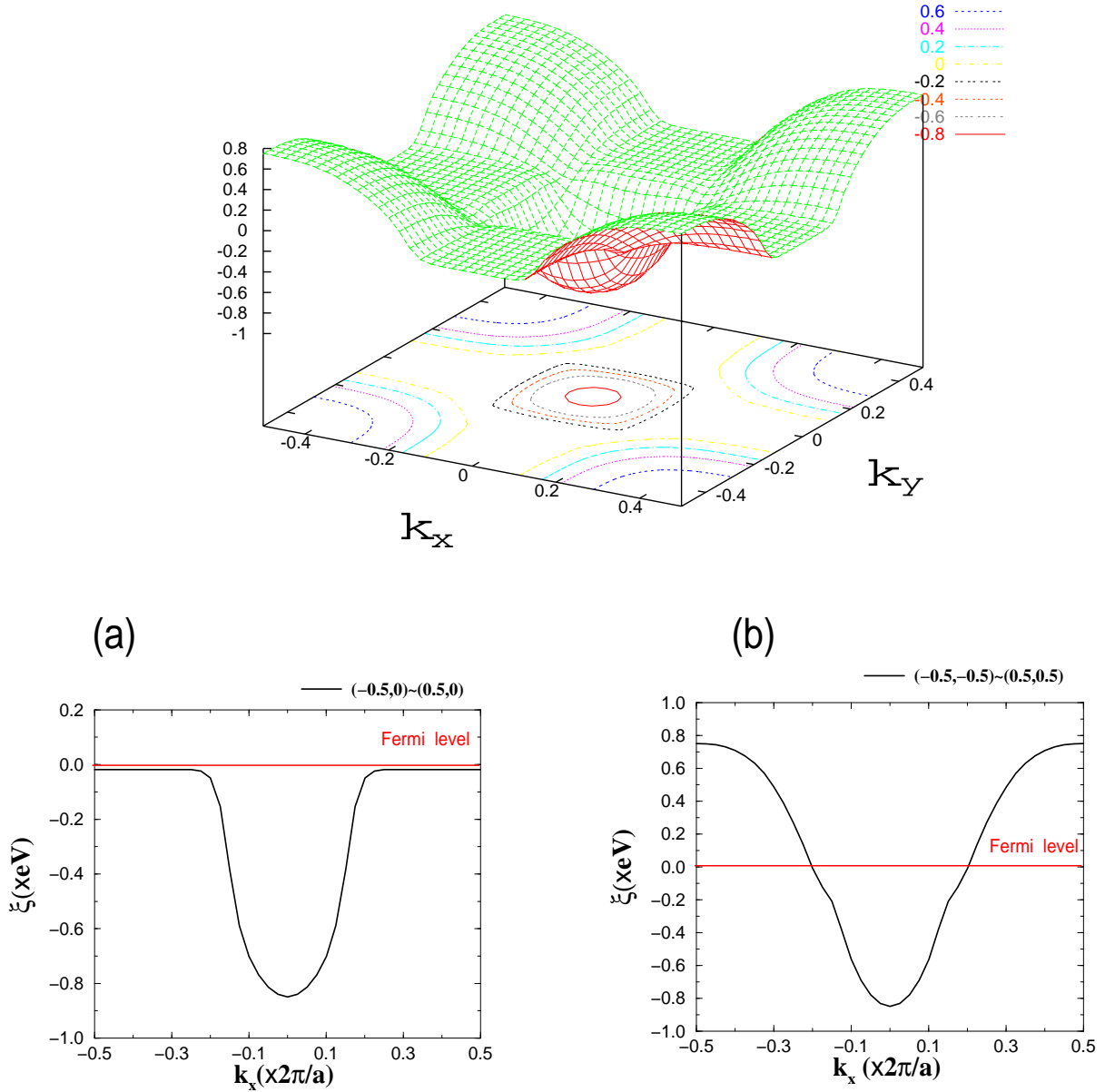} }
\caption{ An typical example of electronic band structure for calculating Tc
and the superconducting order parameter. The parameters required for giving
the flat band region as seen in this figure are $k_{L} = 0.16$, $k_{T} =
0.15 $, and $\protect\lambda_L(\protect\lambda_{T}) = 0.2$ in unit of $2%
\protect\pi/a$, and the parameter for the depth of the flat region $\protect%
\omega_g = 0.018eV$. The $z$-axis indicates the energy of quasi-particles in
units of 1eV. (a)The cut of the band structure from $(-0.5,0)$ to $(0.5,0)$.
The Fermi level is displayed. (b) The cut from $(-0.5,-0.5)$ to $(0.5,0.5)$.}
\end{figure}
\bigskip

\subsection{ Dielectric Function}

Recent inelastic neutron scattering measurements on the optimally doped YBCO
show that the frequencies of the LO and TO phonons of $Cu-O$ bond-stretching
LO mode are inverted in the region of $q_{x}=0.25$ to $0.5$ \cite%
{McQ1,Egami2,Egami3,Chung1}, suggesting that $\epsilon _{el}^{\prime}(q_{x},\omega
^{\ast })$ in this region is negative according to Eq.(7) and the LO phonons
are overscreened by charge fluctuations. It was also observed that the
phonon dispersions are anisotropic along the $a-$ and $b-$axes. In this
paper, however, we neglect this anisotropy for the sake of simplicity, and
assume the tetragonal symmetry of the crystal skipping details of the
dispersion. The effect of the anisotropy will be discussed elsewhere \cite%
{Machida}.

\begin{figure}[tbp]
\centerline{\epsfxsize=14.0cm \epsfysize=12.0cm 
\epsfbox{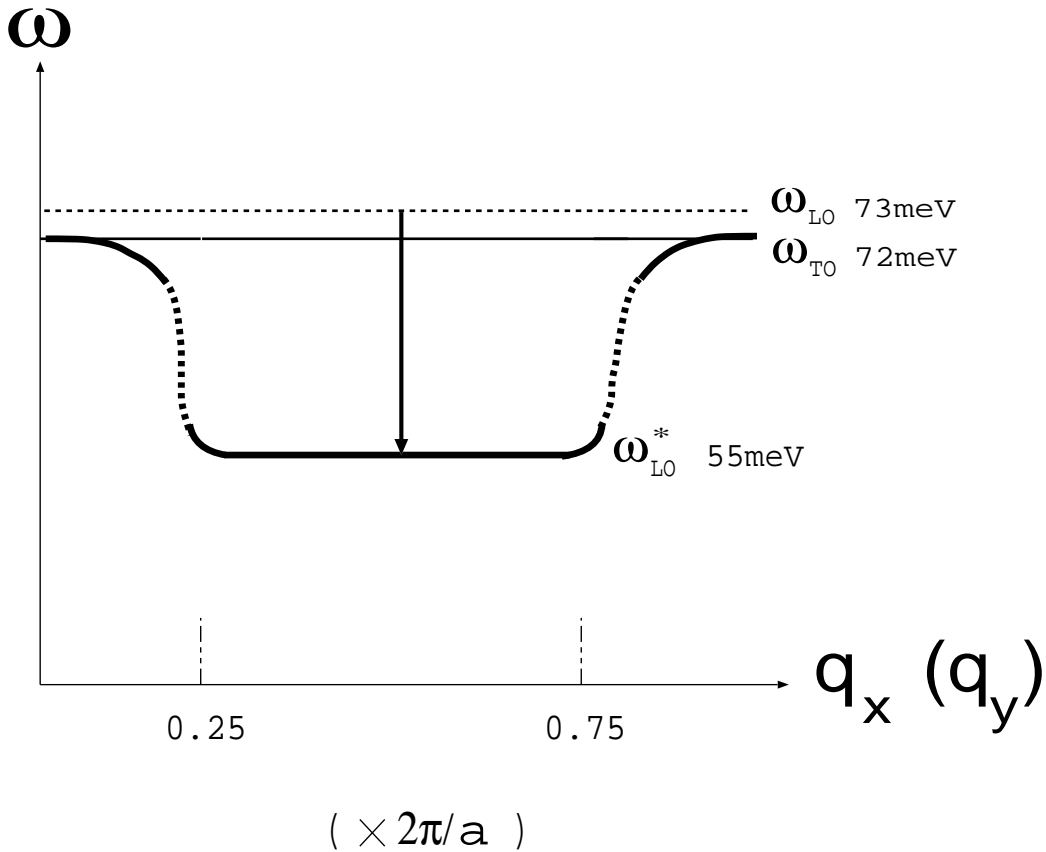} }
\caption{The schematic view of the difference in the energy levels of the
breathing LO and TO phonons between un-doped and doped cuprates for YBCO.
In the undoped state, the LO and TO branches are dispersionless, and are at 73 and
72 meV, respectively, while in the doped state the LO level goes down to
55meV in the range from 0.25 to 0.75 in $q_{x}$ and $q_{y}$ directions
although the TO level still remains dispersionless. The $q_{x}(q_{y})$ is
measured in units of $\protect\pi/a $}.
\end{figure}

\bigskip \noindent We then assume that the dielectric anomalies occur in the
shaded regions as shown in Fig. 3. We divide the ${\bf q}$ space into two
regions, Region II is the shaded area in Fig. 3, where anomalous dielectric
behavior is observed, and Region I which is outside Region II with normal
dielectric behavior. In Region I we assume the Thomas-Fermi type of
screenings of the Coulomb interaction due to quasi-particles, since the
plasma frequency is much higher than $kT_{c}$ and the frequency of the
optical phonons. Then, the inverse of the dielectric function in Region I is
expressed as

\begin{equation}
\frac{1}{\epsilon _{el}({\bf q},\omega )}\approx \frac{1}{\epsilon _{el}(%
{\bf q},0)}=\frac{q^{2}}{q^{2}+q_{TF}^{2}},
\end{equation}%
where $q_{TF}$ is the Thomas-Fermi wave number and its value is an order of
unity in the units of the reciprocal lattice constant for the optimally
doped cuprate superconductors \cite{Homes}. In this region, we note that the
repulsive interaction by the spin fluctuation scattering is located at
(0.5,0.5) \cite{Scalapino,Pines1,Moriya1}, but in the present paper this
interaction is neglected in order to concentrate on the dielectric charge
channel. The repulsive interaction actually exists, but for the moment we
presume that the main pairing contribution arises from the dielectric charge
channel. In Region II, we assume the excitation frequency of the electronic
charge fluctuations is much lower than the plasma frequency for ${\bf q}$ in the region.
The existence of such a low energy charge fluctuation due to anomalous 
electronic structures has been suggested in \cite{Pashitskii}. 
The existence of the low frequency charge
excitation is a necessary condition for the appearance of the negative
dielectric function. Then, in Region II, for the inverse of the dielectric
function for positive low frequencies, we assume the following one pole approximation

\begin{equation}
\frac{1}{\epsilon _{el}({\bf q},\omega )}\approx A(\frac{1}{\omega -\omega
_{0}+i\gamma }),
\end{equation}%
where $A$, $\omega _{0}$, and $\gamma $ are respectively the amplitude, the
frequency of the charge fluctuation density maximum, and the damping
constant. From Eq.(28), A is given by

\begin{equation}
A=\frac{\omega -\omega _{0}}{\epsilon _{el}^{\prime}({\bf q},\omega )}
\end{equation}%
Using Eq.(7) and the experimental values of the frequencies shown in Fig. 2
we estimate $\epsilon_{el}^{\prime}({\bf q},\omega_{LO}^{\ast })$ to be -0.068. The value
of $\omega _{0}$ is much lower than the plasma frequency $(\hbar \omega
_{p}=0.8eV)$ measured at small ${\bf q}$ by reflectivity. Inserting the
value of $\omega _{LO}^{\ast }$ for $\omega $ in Eq.(29), we obtain $A$ to be
from 132 meV if $\hbar \omega _{0}$ is assumed to be 64meV. 
The value of $A$ linearly increases with $\omega _{0}$.
However, we note that Eq.(28) is valid only in the vicinity of the low
charge excitation energy $\hbar \omega _{0}$. Using the value of $A$ and 
$\epsilon _{el}^{\prime\prime}({\bf q},\omega_{LO}^{\ast})$ calculated from 
Eq.(22), we obtain $\gamma$ 
to be $ \sim$ 1 meV. 

\begin{figure}[tbp]
\centerline{\epsfxsize=12.0cm \epsfysize=12.0cm 
\epsfbox{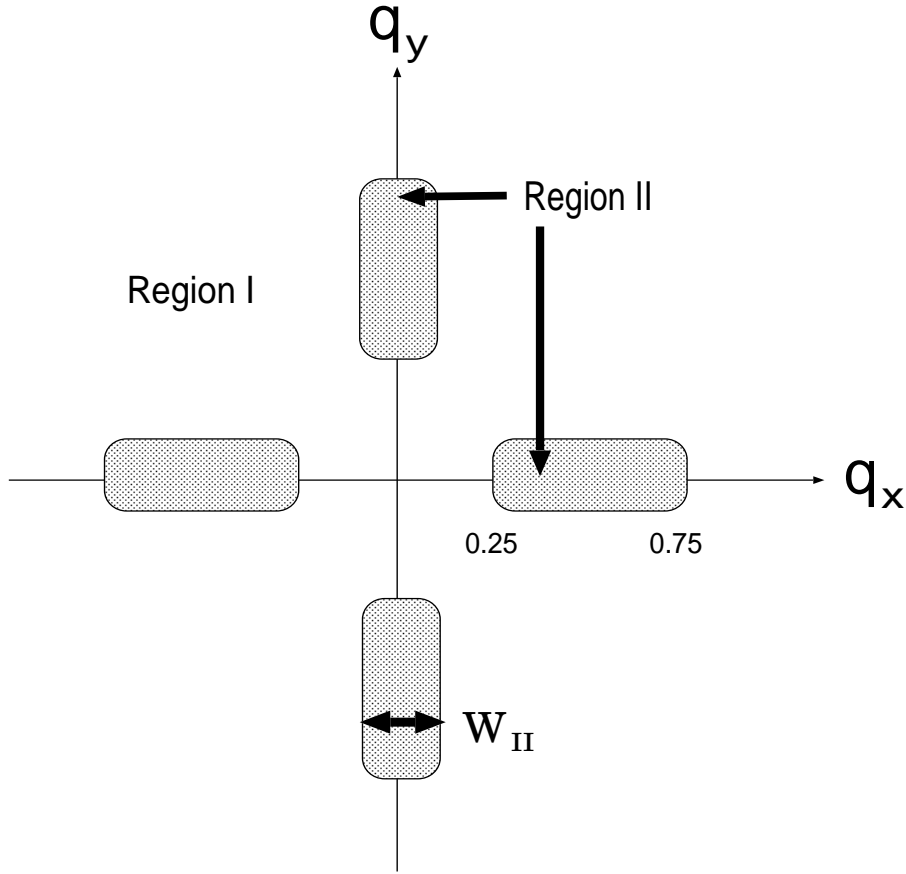} }
\caption{The schematic view of the division into the region I and the region
II (the hatched region) in the space of the scattering wave number between
two quasi-particles. In the region I, the Coulomb interaction is screened as
in a usual metal, while anomalous screening occurs in the region II due to
strong correlation effects. Thus, for the dielectric function, the
expression for Thomas-Fermi type of screening is used in the region I, while
the one obtained from the experimental result (neutron scattering) is used
for the region II. }
\end{figure}
\bigskip

\section{Superconducting Gap Symmetry and $T_c$}

In the previous section, we constructed the model electronic band structure
and derived the dielectric function based on the recent experimental
results. In this section, we determine the kernel, Eq.(20), using the
experimental values given in section VI and calculate $T_{c}$ and the
superconducting gap function by solving Eq.(18).

Let us first concentrate on the kernel function in Eq.(20). The first term
in the bracket in Eq.(20) represents the static dielectric function. It
gives the repulsive contribution in the region I in Fig. 3, while the term
becomes negative causing the attractive interaction in the region II. The
term in the region II is determined by extrapolating the real part of the
dielectric function Eq.(27) to the zero frequency as 
\begin{equation}
\frac{1}{{\epsilon}^{'}_{el}({\bf q},0)}=-A\frac{1}{\omega _{0}}
\end{equation}%
where Eq.(28) is used. At this moment we do not have exact information on
the value of $\omega _{0}$. However, since strong phonon-induced charge
transfer is expected, the value of $\omega _{0}$ must be close to the value
of $\omega _{LO}^{\ast }$, 55 meV. We thus assume the value of $\omega _{0}$
to be 64 meV. Thus, $1/\epsilon _{el}^{\prime}({\bf q},0)$ is
given to be $\sim$ 2. The second term always vanishes on
the Fermi surface since $\xi _{k}=\xi _{k^{\prime }}=0$ on the surface, but
it gives the repulsive interaction in finite energy ranges from the Fermi
level. It is found from the numerical calculations that the contribution
of the second term dominates over the attractive first and third term when $%
\xi_{k}$ and $\xi _{k^{\prime }}$ are sufficiently far from the Fermi level %
\cite{Machida}. The third term gives the attractive interaction originated
from the optical phonons which is renormalized by the electronic dielectric
function. This term is strongly enhanced due to overscreening in the
region II. Such a strong enhancement in the region II can understood by
rewriting and comparing the third term for region I and II, respectively, as
follows, 
\begin{equation}
\frac{q^{2}}{q^{2}+q_{TF}^{2}}\frac{{\omega _{LO}^{\ast }}^{2}({\bf q}%
)-\omega _{TO}^{2}}{{\omega _{LO}^{\ast }%
}^{2}({\bf q})} \quad \mbox{for the region I},
\end{equation}%
and 
\begin{equation}
\frac{1}{\epsilon_{el}^{\prime}({\bf q}, \omega_{LO}^{\ast})  }
\frac{{\omega _{LO}^{\ast }}^{2}({\bf q})-\omega _{TO}^{2}}
{{%
\omega _{LO}^{\ast }}^{2}({\bf q})}\quad %
\mbox{for the region II},
\end{equation}%
where both $\xi _{k}$ and $\xi _{k}^{\prime }$ in the third term are set to
be zero to compare their contributions on the Fermi surface. If $q$ is set
to be $0.5$ for both equations, the reasonable value of $q_{TF}$ is employed
and the observed values $\omega _{LO}^{\ast }$ in both regions are
substituted, it is found that the contribution in the region II becomes more
than about 100 times larger than that in the region I. The reason is explained as
follows. In the region II, the amplitude of $1/\epsilon _{el}^{\ast}({\bf q}%
,\omega )$ almost diverges negatively at $\omega _{0}$. Therefore, the
amplitude of $1/\epsilon _{el}^{\prime}({\bf q},\omega )$ at $\omega \approx \omega_{LO}
^{\ast }$ still remains very large as shown in Fig. 4. The value of Eq.(32) is estimated to be 10.7. 
On the contrary such a anomalous structure does not exist in the vicinity of 
the phonon frequency in the region I and the amplitude of 
$1/\epsilon _{el}^{\prime}({\bf q},\omega )$
remains normal values as in conventional metals. These different features for
the region I and II are seen in Fig. 4.

\begin{figure}[tbp]
\centerline{\epsfxsize=12.0cm \epsfysize=8.0cm 
\epsfbox{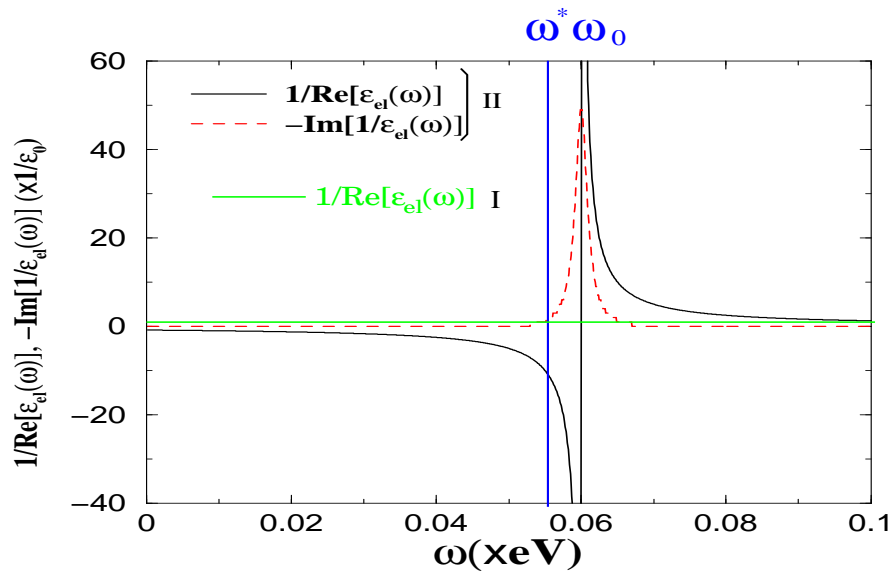} }
\caption{The $\protect\omega $ dependences of $1/\protect\epsilon _{el}^{\prime}(%
\protect\omega )$ in the region I and II and $-Im[1/\protect\epsilon _{el}(%
\protect\omega )]$ in the region II. The solid green line represents $1/
\protect\epsilon _{el}^{\prime}(\protect\omega )$ in the region I, while the solid
black and the dashed red lines stand for 
$1/\protect\epsilon _{el}^{\prime}(\protect\omega )$ and 
$-Im[1/\protect\epsilon _{el}(\protect\omega )]$, respectively. In the region I there 
is no characteristic peak structure
within the frequency range $(0 \approx 0.1 eV)$ because its peak is located
at the high plasma frequency range $(\approx 0.8 eV)$. On the other hand, in
the region II the fluctuation peak lies at $\protect\omega _{0}$ as shown in
this figure.}
\end{figure}
\bigskip \noindent Thus, the electric overscreening effect is found to
enhance the phonon contribution strongly and to give a strong attractive
interaction. In fact, the comparison between the first and the third term
contribution in the region II for the quasi-particles on the Fermi surface
tells us that the third term contribution is 10 times larger than
the first term in this case. This result indicates that the electron-phonon
interaction enhanced by the overscreening could be the main pairing
interaction in the cuprate superconductors.

Next, we actually calculate $T_c$ and the $k$ dependence of the gap function
by solving Eq. (18). For this purpose we use the following technique. We
introduce an eigenvalue equation with an eigenvalue $\lambda(T)$ \cite%
{Tachiki1,Tachiki2}, 
\begin{equation}
\lambda(T) \Phi({\bf k}) = - \sum_{{\bf k^{\prime}}} K({\bf k},{\bf %
k^{\prime}}) \frac{tanh(\xi_{{\bf k^{\prime}}}/2 T)}{2\xi_{{\bf k^{\prime}}}}
\Phi ({\bf k^{\prime}}).
\end{equation}
We numerically calculate $\lambda(T)$ as a function of temperature $T$.
The temperature satisfying $\lambda (T) = 1$ corresponds to the
superconducting transition temperature $T_c$. In addition, since $\Phi(k)$
corresponds to the superconducting gap function, the $k$-dependence of $%
\Phi(k)$ gives the superconducting gap symmetry. However, it should be noted
that the amplitude of the gap function is meaningless since the equation is
a linearized equation which is valid only in the close vicinity of $T_c$.

\begin{figure}[tbp]
\centerline{\epsfxsize=14.0cm \epsfysize=13.0cm 
\epsfbox{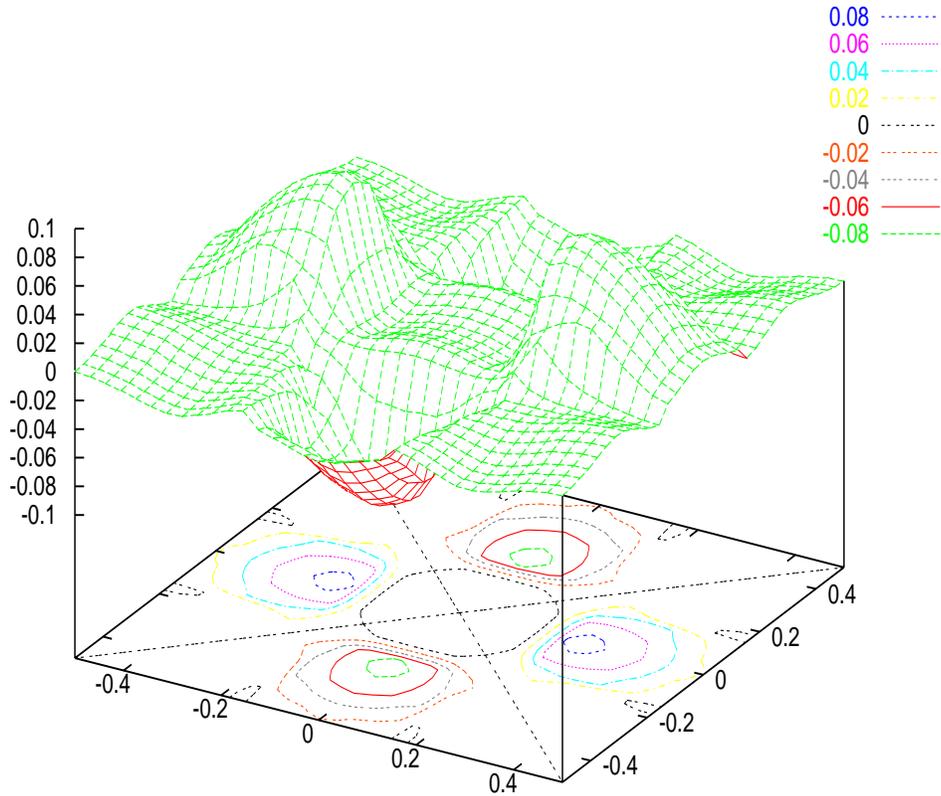} }
\caption{A typical exmaple of the superconducting gap function in the wave
number space. The $x$ and $y$ directions indicate $k_x$ and $k_y$, i.e.,
quasi-particle wave number, respectively, and the $z$-axis gives the
amplitude of the gap function. In the numerical calculation, the parameters $%
\protect\omega_g$, $\protect\omega_0$, $K_L$, $K_T$, and $\protect\lambda_L$
are set to be 13meV, 64meV, 0.16, 0.15, and 0.2, respectively. The
used quasi-particle band structure is displayed in Fig. 1. }
\end{figure}
\bigskip

A typical example of the superconducting gap function is shown in Fig. 5,
where we use 13meV for $\omega _{g}$ in Eq.(23) and 64meV for $\omega _{0}$, 
and the other control parameters are the same as those employed
to depict the band structure shown in Fig. 1 which corresponds to the
optimally doped case. The calculation in this case shows that $T_c$ is about
180K. As seen in Fig. 5 it is found that the gap function has $%
d_{k_{x}^{2}-k_{y}^{2}}$ wave symmetry. The $d_{k_{x}^{2}-k_{y}^{2}}$ symmetry 
is possible under not only the spin fluctuation mediating pairing but also 
other anisotropic pairing interactions \cite{Sherman}.   
 We note that the gap function has
nodes while it shows finite amplitudes even at the center of the flat region
in the electronic structure. This means that the attraction is so strong
that the gap function does not vanish even in regions except for the Fermi
surface. This fact is consistent with the experimental results in ARPES for
the optimally doped case showing high-$T_{c}$ \cite{ARPES}. Here, let us
explain why the $d_{k_{x}^{2}-k_{y}^{2}}$ symmetry appears in this case. The
gap amplitude is generally dependent on the number density of the
interacting quasi-particles and the strength of attraction. Since the number
density of the quasi-particles in the flat region is much larger than other
regions, the gap function is expected to grow considerably in the flat
region. A pair of quasi-particles on the opposite sides of the same flat
 regions along the $k_{x}(k_{y})$ axis are attracted, while the interaction is
 repulsive for a pair of quasi-particles belonging to the flat regions that are rotated 
by $90^{\circ }$ relative to each other. Thus, it is found that the gap amplitudes on the different 
axes should show different signs, resulting in the $d$-symmetry.

\begin{figure}[tbp]
\centerline{\epsfxsize=12.0cm \epsfysize=8.0cm 
\epsfbox{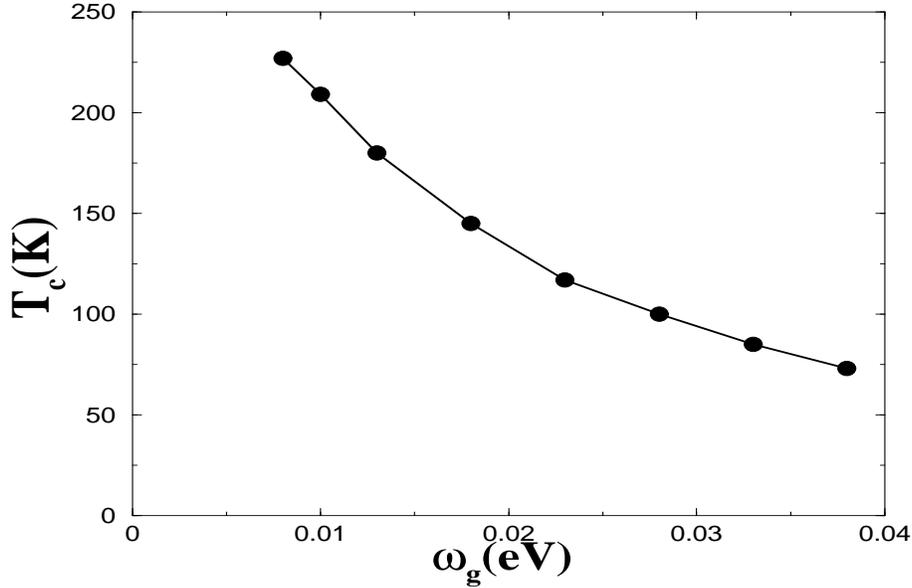} }
\caption{The $\protect\omega_g$(eV) dependence of T$_c$(eV). The other
paramters are the same as ones used in obtaining Fig. 5. }
\end{figure}
\bigskip \noindent 
As seen in Fig. 6, we found that the value of $T_c$ is nearly inversely related to the depth of the flat region. This tendency is consistent with the
experimental results of ARPES \cite{ARPES}, that the depth of the flat region increases 
with increasing doping, and crosses the Fermi level in the overdoped range of composition. 
At the optimum doping, the flat region shows a minimum depth. On the other hand, the value of 
$T_c$ is dependent also on the width $W_{II}$ of Region II, increasing almost linearly with the width. Our calculations show that $T_c$ exceeds 300K when the value of $W_{II}$ approaches to 0.4 with fixing $\omega_g $ to be 13meV, corresponding to the optimum doping.  
It is quite encouraging that such a high value of $T_c$ can be achieved with the present model.

\bigskip

\section{Discussions and Conclusion}

In this paper the mechanism of high $T_c$ superconductivity based upon
overscreening of phonons \cite{Tachiki1,Tachiki2} was developed further
using the results of two recent experimental observations. One is the
neutron scattering measurements showing that the frequencies of a bond
stretching LO phonons in hole-doped superconducting cuprate oxides being
strongly softened in the wave number regions around the Brillouin zone
boundaries along the $k_{x}$ and $k_{y}$ and that the frequency of the LO 
phonon is much lower than that of the TO phonons in these regions 
\cite{McQ1,Egami2,Egami3,Chung1}.
This phenomenon is explained in terms of the overscreening of the ionic
polarization associated with the LO phonon by phonon-induced charge transfer
that results in the negative electronic dielectric function. Another one is
the angle-resolved photoemission measurement indicating that in the
optimally hole-doped cuprates the flat energy band appears just below the
Fermi level around the X (M) points in the Brillouin zone \cite{ARPES,Shen2}%
. Using the pair interaction derived from the neutron scattering results and
the band structure mentioned above, we set up the Eliashberg equation \cite%
{Tachiki1,Tachiki2}. Solving the equation, we observed that the symmetry of
the superconducting order parameter is of $d_{k_{x}^{2}-k_{y}^{2}}$ and the
transition temperature is in excess of 200K. 

The most dominant contribution to the superconducting pairing in the kernel of the Eliashberg equation originates from the phonon contribution that is strongly enhanced by the overscreening effect of the Coulomb interaction. Up to now, several phonon mechanisms leading to d-wave superconducting gap function 
due to the anisotropy of the electronic structure have been suggested \cite{anisophonon}. In contrast, the present theory relies upon a mechanism of phonon overscreening resulting in the negative electronic dielectric function and strong enhancement of its attractive contribution. 
As the most of the phonon mechanisms, the present theory predicts that 
the isotope effect \cite{Isotope} appears in the cuprate superconductors with a relatively low T$_{c}$ such as LSCO, but this effect may not be visible in the cuprates with a high T$_{c}$ such as YBCO and Bi-2212, as observed by experiments. This result comes from the fact that as the superconducting transition temperature increases, the pure phonon contribution for the pairing
diminishes and the electronic contribution increases instead. This tendency is
consistent with the experimental results \cite{Isotope}.

The softening of the LO phonon near the Brillouin zone boundary was observed
by neutron scattering even in the superconducting Ba$_{0.6}$K$_{0.4}$BiO$%
_{3} $ with negligible spin fluctuation \cite{BKBO}, and non-superconducting
La$_{1.69}$Sr$_{0.31}$NiO$_{4}$ \cite{LNiO} and La$_{0.7}$Sr$_{0.3}$MnO$_{3}$
\cite{LMnO}. It appears that the overscreening effect is a common nature in
hole-doped transition metal oxide systems. This phenomenon may be explained
in terms of the "negative Born effective charge", which will be further
elaborated later \cite{Piekarz}. This concept nicely explains the phonon
softening observed for the Peiels-Hubbard Hamiltonian in one dimensional
Cu-O chain model \cite{Ishihara1,Petrov1}. 

In the cuprates the bare bandwidth, calculated for instance with the LDA, is
of the order of 1 eV. Band-narrowing is brought about by spins that almost localize charges and create the extended saddle point in the electronic band structure. 
In the present paper this effect was considered phenomenologically 
by using the renormalized band structure. However, a more satisfactory theory should 
include the quantum effect of spins more explicitly. Using the $p-d$ model hole doping was found to create in-gap states in the Hubbard gap just at the Fermi level, with a dominant oxygen $p$-character \cite{Matsumoto}. A better model band structure with these features needs to be developed. A problem here is that 
we have much less information about the unoccupied states compared 
to the occupied states, since the photoemission experiments provide 
information only for the latter, while the inverse-photoemission technique, which should give the information on the former, has much less resolution.

As we mentioned above in the present mechanism LO phonons excite holes from
the local oxygen $p$-state to the Cu $d$-state. The results of inelastic
neutron scattering measurements on YBa$_{2}$Cu$_{3}$O$_{6.95}$ \cite{Chung2}
suggest the final state of this transition could be the $d_{z^2}$ level of
Cu, since strong mixing of the in-plane Cu-O bond-stretching mode and the
apical oxygen mode (62 meV) was observed. This makes the two-band phononic
model \cite{Annette} relevant in the present context. In addition there is a
possibility of the transition from an oxygen in the $x$-direction to another
oxygen in the $y$-direction contributing to the $e-p$ coupling \cite{Mihail}%
. In the present calculation just one band was assumed for the sake of
simplicity, but the reality is likely to be more complex.

Another point that has not been included in the present paper, and has to be
addressed in future publications, is the question of the in-plane anisotropy and
the stripe fluctuations. Both the inelastic neutron scattering measurement
of the phonon dispersion and the photoemission measurement detected
surprisingly strong anisotropy in the CuO$_{2}$ plane \cite%
{Egami2,Egami3,Endoh,Shen2}. This anisotropy most likely is related to
the spin-charge phase separation in the form of stripes \cite{Tranq1}.
 While static stripes apparently compete against superconductivity \cite{Tranq2}, it is reasonable 
to assume that the propensity for stripe formation remains strong even 
in the superconducting phase \cite{Yamada1,McQ1,Egami2,Egami3}. The
presence of stripes or stripe fluctuations will produce further narrowing of 
the band and spin-charge interference phenomena, which should enhance the
superconductivity. The possibility of such enhancement by confinement, in a
more general sense, was emphasized by Phillips \cite{Phillips}. However, since 
we do not have sufficient information to formulate these phenomena 
into a rigorous theory, in the present paper we have not explicitly taken them into account. Further experimental as well as theoretical researches are needed to fully account for the spin-charge synergy.

In the present mechanism the role of spins is merely to bring down the
energy scale of holes to the level of phonons, so that the vibronic
resonance can take place. We do not, however, exclude the possibility that
the magnetic mechanism also contributes to pairing, at least in some ranges of
composition. Even in such a case the phonon mechanism does not compete
against the magnetic mechanism because of the $d$-symmetry. It is most
likely that the relative importance of the phononic and magnetic mechanisms
depends upon charge density; magnetic mechanism could be more important in
the underdoped region, while the phononic mechanism may dominate the optimum
and overdoped region.

In the present paper we sacrificed some details as discussed above and
greatly simplified the model. Furthermore it is well known that the calculation of $T_{c}$ in the
Eliashberg theory strongly depends upon details of the parameters. The
purpose of this work, therefore, is not to try to describe the
superconductivity in the cuprates very accurately, but to
demonstrate the possibility of the phononic mechanism of high-$T_{c}$
superconductivity based upon the overscreening phenomenon. The estimated value
of $T_c$ is even higher than the experimental values over wide ranges of parameters. 
This is in part due to the fact that the
calculation of Tc is essentially based on a mean field theory 
and does not include the superconducting fluctuation effect. 
Inclusion of the fluctuations should significantly decrease $T_c$, since the pairing force is strong and the system has a two dimensional nature. 
The incoherent superconducting fluctuations will be observed as the pseudo gap above $T_{c}$.

While there are a number of limitations in the present model because of the simplifications and the phenomenological nature, the result described here strongly suggests that the phononic mechanism 
has to be taken seriously, and careful studies are warranted. 
We will address some of the problems that are neglected here in future publications. 

\section{Acknowledgements}

We are indebted to K. A. M\"{u}ller, A. A. Abrikosov, J. C. Phillips, A. Bussmann-Holder, S. Sachdev, N. Nagaosa, D. Mihailovic, V. V. Kabanov, S. Takahashi, M. R. Norman, N. Hamada, and M. Arai for illuminating discussions. We wish to thank Y. Endoh and J. M. Tranquada for sharing their unpublished experimental results. Two authors (M. T. and M. M.) were supported by CREST (JST) project. The other (T. E.) was supported by the National Science Foundation grant DMR01-02565. A part of this work was done in Argonne National
Laboratory. M. T. and M. M. thank G. Crabtree for his support and M. M.
especially thanks A. E. Koshelev for his support and valuable discussions.
M. M. also thanks T. Imamura for offering of a numerical diagonalization program 
and the stuff members of CCSE in JAERI for their computational assistance.

\end{document}